\documentclass[12pt]{article}

\usepackage{amsmath,amssymb,amsfonts,amsbsy}
\usepackage{cite}
\usepackage{graphicx}
\usepackage{wrapfig}
\usepackage{epsfig}

\usepackage{bbm} 
\usepackage{bm} 
\usepackage{color}                                                       %
\usepackage{dsfont} 
\usepackage{latexsym} 
\usepackage{lscape} 
\usepackage{mathrsfs} 
\usepackage{morefloats} 
\usepackage{floatflt} 
\usepackage{slashed} 
\usepackage{psfrag}

\DeclareFontFamily{OT1}{mygreek}{}%
\DeclareFontShape{OT1}{mygreek}{m}{n}{<->omsegr}{}%
\DeclareFontShape{OT1}{mygreek}{b}{n}{<->omsegrb}{}%
\DeclareFontShape{OT1}{mygreek}{m}{it}{<->omsegri}{}%
\DeclareFontShape{OT1}{mygreek}{bx}{n}{<->sub * mygreek/b/n}{}%
\DeclareFontShape{OT1}{mygreek}{m}{sl}{<->sub * mygreek/m/it}{}%
\DeclareSymbolFont{Greekrm}{OT1}{mygreek}{m}{n}
\DeclareSymbolFont{Greekbf}{OT1}{mygreek}{b}{n}
\DeclareSymbolFont{Greekit}{OT1}{mygreek}{m}{it}
\DeclareMathSymbol{\omegab}{\mathalpha}{Greekbf}{119}

\begin{document}
\addcontentsline{toc}{subsection}{{Magnetic Polarizability of Diquarks in Baryons}\\
{\it P. Filip}}
\graphicspath{{exper/yourname/}}

\setcounter{section}{0}
\setcounter{subsection}{0}
\setcounter{equation}{0}
\setcounter{figure}{0}
\setcounter{footnote}{0}
\setcounter{table}{0}

\begin{center}
\textbf{MAGNETIC POLARIZABILITY OF DIQUARKS IN BARYONS}

\vspace{5mm}

{P.~Filip}

\vspace{5mm}

\begin{small}
   \emph{Institute of Physics, Slovak Academy of Sciences } \\
   \emph{D\'ubravsk\'a cesta 9, Bratislava 845 11, Slovakia} \\
   \emph{E-mail: Peter.Filip@savba.sk}
\end{small}
\end{center}

\vspace{0.0mm} 

\begin{abstract}
  We study the response of diquark wave function in $\Lambda$-type baryons to strong magnetic fields. 
It is found that quantum state of $J$=0 diquark ($ud$) in the magnetic field changes
due to magnetic polarizability, and constituent quarks in ($ud$) diquark become polarized. 
The phenomenon influences polarized quark distribution functions $\Delta u(x)$ and  $\Delta d(x)$, 
which therefore may be sensitive to the internal electromagnetic fields in hypernuclei.  
We also speculate, that strange quark polarization in nucleon
may originate from the interaction of virtual $s\bar s$ quark pairs with the intrinsic magnetic field of nucleon 
$B\approx 10^{13}$T. 

\end{abstract}

\vspace{7.2mm}

\section{Introduction}

It has been suggested many years ago \cite{MGM_Zweig}, that baryons and mesons 
contain fractionally charged fermions - constituent quarks. According to Dirac equation,
magnetic moment of charged particles with spin $s=1/2$ is $\mu$=$\hbar Q/2m^*\!$, and therefore,
constituent quarks should have magnetic moments. For baryons this concept works surprisingly well, 
and measured magnetic moments of hyperons
$\Omega^-,\Xi^0, \Xi^-,\Sigma^+,\Sigma^-,\Lambda^0$, proton and neutron,
can be understood as originating from the 
magnetic moments $\mu_u$=1.85$\mu_N$, $\mu_d$=-0.97$\mu_N$, $\mu_s$=-0.61$\mu_N$ 
of quarks with constituent masses $m^*_u, m^*_d\,$$\approx$ 330MeV and 
$m^*_s$=510MeV.

Consequently, open-flavor vector mesons should also have magnetic moments.
For example, $K^{*+}$ meson (bound state of $u,\bar s$ quarks with parallel spins) may be expected to have
magnetic moment $\mu_{K^{+*}} = |\mu_u| + |\mu_s| = 2.5\mu_N$
(here $\mu_N$=3.1$\cdot$10$^{-14}$MeV/T).

The response of pseudoscalar mesons and scalar diquarks to external magnetic fields 
can be understood using the analogy of $(qq')$ bound states with muonium ($e^-\mu^+$) and 
positronium ($e^-e^+$). Similarly to singlet ($J$=0) ground state of positronium or muonium,
mesons 
$\eta_c$, $\eta_b$, $\eta'$, $\pi$, $K$, $D$, $B$ should have zero magnetic moment \cite{HS_2013}. 
In the magnetic field however, due to magnetic polarizability of pseudoscalar mesons,
induced magnetic moment $\tilde \mu[B]$ is expected to appear \cite{My_SPIN13}, 
due to partial polarization of $q\bar q$ pair in $J$=0 quantum state.
If the analogy with positronium behavior \cite{HalpernRich} is indeed correct, wave function 
$(\uparrow\downarrow + \downarrow\uparrow )/\sqrt{2}$ 
of ($m_z$=0) substate of vector mesons can
acquire the admixture of pseudoscalar state
$(\uparrow\downarrow - \downarrow\uparrow )/\sqrt{2}$ in the magnetic field, and
quenching \cite{QuenchingEE} of 
$\Psi(c\bar c)$, $\Upsilon(b\bar b)$ and $\varphi(s\bar s)$ meson decays may occur \cite{CPOD_2013}
in static external fields $B\approx 10^{14}-10^{15}$T. 

Internal spin structure of scalar diquarks \cite{RMP_diquarks} in $\Lambda$-type baryons 
resembles quantum state of pseudoscalar mesons: 
$(\uparrow\downarrow - \downarrow\uparrow )/\sqrt{2}$. In strong magnetic field, a
superposition of ($J$=0) diquark with its excited state ($J$=1, $m_z$=0) can take place.
In this contribution we discuss the magnetic polarizability of
diquarks in baryons due to  
fields $B\approx 10^{11}$-$10^{14}$T.

\section{Spin structure of $\Lambda$ baryons} 

\begin{wrapfigure}[10]{R}{55mm}
  \centering 
  \vspace*{-9.9mm} 
  \includegraphics[width=50mm]{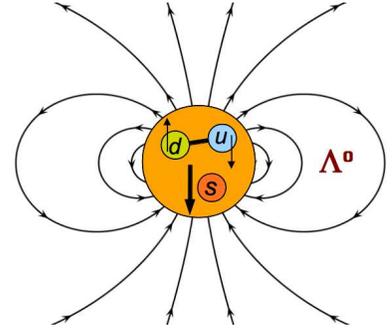}
  \caption{\footnotesize Scalar diquark ($ud$) in $\Lambda^0$ hyperon. 
   The field is generated by magnetic moment $\mu_s$ of $s$-quark.}
  \label{yourname_fig1}
\end{wrapfigure}
Internal spin structure of $\Lambda^0_{1/2}(1116)$ baryon differs from that of proton, neutron
and other  spin 1/2 hyperons\cite{Franklin_PR172}. 
Typical $s=1/2$ baryon contains two quarks (diquark) in ($J$=1) triplet state accompanied 
with the third quark, as described by naive $SU(6)$  
function $\varPsi^T_{1/2}$ in Eq.(1).  
One may directly guess that ground state wave function  
of proton ($uud$) is similar to $\Sigma^+$ ($uus$), 
since both they contain quarks ($uu$) accompanied by third $d$ or $s$ quark. 
Almost equal masses of  $\Sigma^+, \Sigma^-, \Sigma^0$ hyperons
then suggest, that their constituent quantum spin structure is similar,
(given by $\varPsi^T_{1/2}$) as in the case of proton and neutron.
\begin{equation}
\varPsi^S_{1/2}= 
\frac{(\uparrow\downarrow-\downarrow\uparrow)\uparrow}
        {\sqrt{2}}
\qquad
\varPsi_{1/2}^T = 
\frac{\sqrt{2} \uparrow\uparrow\downarrow - (\uparrow\downarrow+\downarrow\uparrow)\uparrow/\sqrt{2}}
        {\sqrt{3}}
\qquad 
\varPsi_{3/2}= (\uparrow\uparrow\uparrow)
\label{Eq1}
\end{equation}

However, constituent quarks ($uds$) of $\Sigma^0_{1/2}$ hyperon can enter a lower-energy quantum state $\varPsi^S_{1/2}$, 
with different configuration of quark spins. Such state is observed experimentally as $\Lambda^0_{1/2}$(1116). 
Mass difference ($\delta M$=77MeV) between $\Sigma^0(1193)$ and $\Lambda^0$ baryon comes
from different interaction energy of constituent quark color-magnetic moments.

Quantum structure of  $\Lambda^0$ hyperon thus contains scalar 
($J$=0) diquark accompanied with the third 
quark, which is then responsible for the spin 
of such baryon. A question, 
which quarks enter the scalar ($J$=0) diquark state in flavor-degenerate baryons of type 
$\Lambda^0(uds)$, $\Xi^+_c(usc)$ or $\Omega^0_{cb}(scb)$ has been
discussed already by Franklin et al. \cite{Franklin_PRD24}.
The conclusion was that two quarks with similar 
masses form a scalar diquark state, with small admixture of other diquark flavor configurations.
For $\Lambda^0(uds), \Lambda^+_c(udc), \Lambda^0_b(udb)$
this means that scalar diquark ($ud$) is accompanied with heavier $s$, $c$ or $b$ quark.
 
If all three constituent quark spins are oriented in parallel, 
baryon has spin $s$=3/2, which corresponds to experimentally observed $\Omega$ hyperon
and $\Delta$, $\Sigma^*$ and $\Xi^*$ resonances. Spin wave 
function $\varPsi_{3/2}$ of such baryons is shown Eq.(\ref{Eq1}).

\section{Internal hyperfine magnetic fields in baryons}
Within the framework of MIT bag model \cite{MIT_bag}, constituent quarks are bound together
in a small ($R\approx 1$fm) volume, which contains strong gluon fields and also virtual partons.
Constituent quarks are the source of magnetic dipole and electric fields, which are not screened
by the external vacuum. 
In a simplified picture of $\Lambda^0$ hyperon as purely ($ud$)-$s$ state, 
the measured magnetic moment $\mu_\Lambda = -0.613\mu_N$ 
is to be generated by $s$-quark: $\mu_\Lambda = \mu_s$, because diquarks in quantum 
state $\varPsi^S = (\uparrow\downarrow-\downarrow\uparrow)\sqrt{2}$
(as well as pseudoscalar mesons) should have zero magnetic moment. 

However, the above said is not completely true. Magnetic dipole field lines, which constitute
the dipole field of $\Lambda^0$ hyperon are contained in (penetrating) the "bag" volume of baryon (see Fig.1).
Therefore, scalar ($ud$) diquark state, described by the spin wave function $\varPsi^S$
can be altered in the magnetic field, and achieve (due to its magnetic polarizability)
an induced magnetic moment $\tilde \mu[B]$, as discussed for the $\eta$-meson case in \cite{My_SPIN13}. 
In such picture, virtual quark-antiquark pairs and scalar diquarks are swimming in a deconfined QCD medium
(the "bag") containing also gluons and strong magnetic field. 

Let us estimate the strength of hyperfine magnetic field inside baryons:
Since the source of the magnetic dipole field is localized inside the hadronic "bag" volume,
we shall assume, that dipole magnetic moment $\mu$\,=\,$c_1\!\cdot\!\mu_N$ of baryon
comes from the fictious current loop of radius 
$R_B$=$r_o$[$10^{-15}$m] (for proton $c_1$=2.79, for 
$\Lambda$ hyperon $c_1$=$-$0.61). One has
\begin{equation}
\mu = I \cdot S = I \cdot \pi R_B^2 \qquad \longrightarrow \qquad I = 
(c_1/\pi r_o^2)\mu_N 10^{30} \approx  5 (c_1/\pi r_o^2) 10^3 A
\end{equation}
using $\mu_N = 5\cdot 10^{-27}$ J/T. 
Magnetic field $B_{int}$ at the center of such current loop is
\begin{equation}
B_{int} = \mu_o I / 2 R_B 
\qquad \longrightarrow \qquad B_{int} \approx (2 c_1/r_o^3) 10^{12}\, T \,\, ,
\label{magFieldLoop}
\end{equation}
if magnetic permeability $\mu_o = 4\pi\cdot 10^{-7}$NA$^{-2}$ of vacuum is used. 
For $\Lambda^0$ hyperon we then obtain internal magnetic field
$B_{int}^\Lambda \approx 4\cdot 10^{12}$T (assuming $r_o$=0.67 [fm]), and  for proton 
$B_{int}^p \approx 10^{13}$ Tesla, assuming fictious current loop 
radius $r_o=0.82$ [fm]. 

\section{Scalar diquarks in the magnetic field}
External and intrinsic magnetic field of baryons can influence quantum state of scalar diquarks via interaction term: $H_{int}$=$-\vec\mu_q\!\cdot\!\vec B$. Similarly to the case of Positronium
and Muonium, spin-singlet state $\tilde \varPsi^S[B]$ becomes a quantum superposition of
triplet and singlet states \cite{HalpernRich}, and induced magnetic moment \cite{CPOD_2013} 
of scalar ($ud$) diquark appears
\begin{equation}
\tilde \varPsi^S[B] = \frac{c_\alpha-s_\alpha}{\sqrt{2}} | \uparrow \downarrow \rangle 
-  \frac{c_\alpha+s_\alpha}{\sqrt{2}}| \downarrow\uparrow \rangle
; \qquad
\langle \tilde \varPsi^S |\hat{\mu} |\,\tilde\varPsi^S \rangle =  (|\mu_{u}|\!+\!|\mu_{d}|) \sin(2\alpha)
= \Delta\mu
\label{Eq4}
\end{equation}
where $s_\alpha\!=\!\sin(\alpha)\!= y/\sqrt{1+y^2}$, $c_\alpha\!=\cos(\alpha)\!=\!\sqrt{1-s_\alpha^2}$, and 
$y = x/(1+\sqrt{1+x^2})$ depends on magnetic field $B$
via parameter $x=2(|\mu_{u}|+|\mu_{d}|)B/\Delta E_{hf}$. Hyperfine splitting $\Delta E_{hf}$
is $(M_\Lambda - M_\Sigma)$ = 77MeV for $\Lambda^0$ hyperon, and 
166MeV and 194MeV for $\Lambda^+_c$ and $\Lambda_b$ hyperons.  
In the limit $B\rightarrow \infty$, $\alpha = 45^{\circ}$, and scalar diquark becomes
fully polarized $\tilde \varPsi^S$=$-|\!\downarrow\uparrow \rangle$ in its ($J$=0) state: 
quark magnetic moments become oriented along field $\vec B$ direction,
while their spins are anti-parallel. In such extreme case, polarized quark 
distribution functions $\Delta u(x)$ and $\Delta d(x)$ of $\Lambda^0$ baryon are substantially affected.

Induced magnetic moment $\Delta \mu$ of scalar diquark should
contribute to the magnetic moment of $\Lambda^0$ hyperon, as pointed out already
by Franklin et al.\cite{Franklin_PRD24}. In the limitting case 
$\Delta \mu \rightarrow |\mu_u| + |\mu_d| = 2.8\mu_N$. 
For our intrinsic magnetic field $B = 4\cdot 10^{12}$T in $\Lambda^0$ hyperon: 
$\sin(2\alpha) \approx x = 2(|\mu_{u}|$+$|\mu_{d}|)B/\Delta E_{hf} = 0.0091$ and $\Delta\mu = 0.026\mu_N$,
which is 4\% of $\mu_\Lambda$. Here, we did not take into account the full wave function
$\varPsi^T_{1/2}$ of $\Sigma$ baryon (see Eq.\ref{Eq1}), which contains term $(\uparrow\downarrow+\downarrow\uparrow)\sqrt{2}$ with probability $(1/\sqrt{3})^2$. 
Magnetic polarizability of scalar ($us$) and ($ds$) diquarks in $\Xi_c$ 
hyperons 
originates from the same mechanism: the superposition of $\varPsi^S$ with $\varPsi^T$ triplet
state of $\Xi_c'$ hyperons (they correspond to $\Sigma^0$).
Due to different quark magnetic moment orientation relative to quark spin in ($us$) and ($ds$) diquarks, 
magnetic polarizability $\beta_0=2\langle \Psi^S | \hat{\mu}_{ds} | \Psi^T\rangle^2\!/\Delta E_{h\!f}$ 
of ($ds$) diquark is expected to be much smaller compared to ($us$) and ($ud$) diquarks 
(see Eq.10 and Eq.11 in \cite{HS_2013}).

The interaction of color-magnetic dipole moments of quarks induces
additional hyperfine mixing \cite{Franklin_PR172, Franklin_PRD24} 
of wave functions $\varPsi^T$ and $\varPsi^S$, which is
independent from  purely electromagnetic effects we study here.

\section{Virtual $s\bar s$ pairs polarization in nucleon}
Similarly to virtual $e^+e^-$ pairs, which 
contribute to anomalous magnetic moments of electron and muon, 
virtual ($s\bar s$) pairs can influence nucleon properties.
Various experimental results suggest, that 
($s\bar s$) quark pairs in nucleon are polarized: $\Delta s = -0.1\pm 0.02$ \cite{Shaposhznikov}.

Let us assume here, that intrinsic magnetic field $B_{int} \approx 10^{13}$T in nucleon affects 
quantum state of virtual ($s\bar s$) pairs. 
Inside the hadronic bag, without any external fields,
virtual $s\bar s$ pairs would appear in pure $J$=$0^{+-}$ singlet state
$\Psi^S =  (\uparrow\downarrow - \downarrow\uparrow)/\sqrt{2}$,
or in $J$=$1^{--}$ triplet state. 
Due to its smaller energy, pseudoscalar configuration $\Psi^S$ should be more probable.
If internal magnetic field $B_{int}\approx 10^{13}$T in nucleon modifies the wave function 
$\tilde\Psi^S[B]$ of scalar $s\bar s$ pairs
as described by Eq.(\ref{Eq4}), induced magnetic moment 
of $J$=0 ($s\bar s$) pairs appears: $\langle \hat\mu \rangle_{s\bar s} = 2|\mu_s| \sin(2\alpha)$, which 
may contribute to the nucleon magnetic moment. 
At the same time, net polarization of virtual $s$ quarks occurs. 

\section{Conclusions}
We have discussed that quantum state of scalar diquarks in $\Lambda^0$ - type hyperons
can be influenced by internal and external magnetic fields. Our estimate of the intrinsic (hyperfine) magnetic field
for $\Lambda^0$ hyperon is $B_{int} = 4\cdot 10^{12}$T.
We suggest, that  polarized quark distribution functions $\Delta q(x)$ of $\Lambda^0$- type 
hyperons can be modified 
due to polarization of scalar ($ud$) diquark in strong electromagnetic field,
which may be remotely related to EMC effect. We also suggest, that virtual $s\bar s$ pairs in nucleon
are effectively polarized due to the intrinsic magnetic field of nucleon $B_{int} \approx 10^{13}$T.
\newline
\textbf{Acknowledgement}:\! This work is supported by Slovak Grant Agency VEGA (2/0197/14).

\end{document}